\documentclass[%
aps,
rsi,%
amsmath,amssymb,
preprint,
]{revtex4-1}

\usepackage{dcolumn}
\usepackage{import}
\usepackage{graphicx}
\usepackage{xcolor}
\usepackage{amssymb}
\usepackage{geometry}
\usepackage{csquotes}
\usepackage{placeins}

\begin{document}

\title{Crystal polarity discrimination for GaN nanowires on graphene}

\author{Alexander Pavlov} 
\email{a.pavlov@physics.spbstu.ru}
\affiliation{Alferov University (Former St Petersburg Academic University),
                St. Petersburg, 194021, Russian Federation}
\affiliation{Peter the Great St.Petersburg Polytechnic University, 
              Polytechnicheskaya 29, 195251, St.Petersburg Russia}    

\author{Alexey Mozharov}
\affiliation{Alferov University (Former St Petersburg Academic University),
                St. Petersburg, 194021, Russian Federation}
                
\author{Yury Berdnikov}
\affiliation{St. Petersburg State University, Saint Petersburg, 199034, Russian Federation}
   
\author{Camille Barbier}
\affiliation{Centre de Nanosciences et de Nanotechnologies (C2T), 
              Univ. Paris-Saclay, UMR 9001 CNRS, 10 Boulevard Thomas, 
              Gobert, 91120 Palaiseau, France.}

\author{Jean-Christophe Harmand}   
\affiliation{Centre de Nanosciences et de Nanotechnologies (C2T), 
              Univ. Paris-Saclay, UMR 9001 CNRS, 10 Boulevard Thomas, 
              Gobert, 91120 Palaiseau, France.}
                
\author{Maria Tchernycheva}
\affiliation{Centre de Nanosciences et de Nanotechnologies (C2T), 
              Univ. Paris-Saclay, UMR 9001 CNRS, 10 Boulevard Thomas, 
              Gobert, 91120 Palaiseau, France.}

\author{Roman Polozkov} 
\affiliation{ITMO University, Saint Petersburg, 197101, Russian Federation}
\affiliation{Alferov University (Former St Petersburg Academic University),
                St. Petersburg, 194021, Russian Federation}
    
\author{Ivan Mukhin}
\affiliation{Alferov University (Former St Petersburg Academic University),
                St. Petersburg, 194021, Russian Federation}
\affiliation{ITMO University, Saint Petersburg, 197101, Russian Federation}

\begin{abstract}
We present experimental data and computational analysis of the formation of GaN nanowires on graphene virtual substrates. 
We show that GaN nanowires on graphene exhibit nitrogen polarity. 
We employ the DFT-based computational analysis to demonstrate that 
among different possible configurations of Ga and N atoms only the N-polar one is stable. 
We suggest that polarity discrimination occurs due to the dipole interaction 
between the GaN nanocrystal and $\pi$-orbitals of the graphene sheet. 
\end{abstract}

\maketitle

\section{Introduction}

Gallium nitride (GaN) is a wide bandgap semiconductor with unique optical, 
electrical and chemical properties \cite{morkocc2009handbook}. 
It can by synthesized in the form of nanowires (NWs), 
which are intensively investigated as active structures for various 
optoelectronic applications, including light-emitting diodes 
\cite{guo2010catalyst, hersee2009gan}, 
solar cells 
\cite{tian2009single, tang2008vertically, 
mozharov2015numerical, neplokh2016electron, shugurov2019study}, 
photodetectors \cite{rigutti2010ultraviolet, gonzalez2012room}
and piezogenerators \cite{jamond2016piezo, gogneau2016single, lu2018probing}. 
%
%
Generally, GaN NWs are grown on bulk crystalline substrates, such as silicon, 
\cite{guo2010catalyst, calarco2007nucleation, largeau2012n, 
      bolshakov2019effects, fedorov2018droplet, bolshakov2018dopant, gridchin2020selective}
sapphire \cite{hersee2006controlled, wang2006highly, avit2014ultralong}
or diamond \cite{hetzl2016reprint}.
In this case, the substrate imposes to NWs an epitaxial relationship, 
which determines the NW growth direction and sidewall orientation.  
Also, this type of nanostructures can be grown on alternative substrates such 
as metal \cite{calabrese2016molecular} 
or even amorphous materials like glass \cite{kumaresan2016self}. 
%
While vertical NW orientation may be preserved in these cases, the in-plane orientation of 
NWs is arbitrary.

In the last decade, a new class of virtual substrates made of two- dimensional 
(2D) materials has attracted a strong attention \cite{hong2015van}. 
It was observed that the transfer of 2D flakes or sheets to almost any substrate 
can promote the epitaxial growth of both vertically and in-plane oriented NW arrays.
%
This approach, termed "van der Waals epitaxy", either reduces or completely 
eliminates the influence of the host substrate on the crystal properties 
of synthesized nanostructures. 
Recently, graphene flakes have been explored as substrates to grow 
selectively III-V NWs \cite{MNL, munshi2012vertically, gridchin2020selective}. 
Graphene can be considered not only as a virtual substrate for the NW growth, 
but in some cases also as a transparent electrode with simultaneously excellent mechanical and 
electrical contact to NWs’ bases \cite{KoreaPaper}.  
Moreover, due to weak van der Waals interaction with the host substrate, 
graphene with NWs encapsulated into polymer matrix can be exfoliated 
from the rigid substrate providing very promising platform 
for flexible optoelectronic applications. 
Presumably, the contact of NWs to graphene can be preserved after 
matrix exfoliation \cite{KoreaPaper}.

Today, a number of publications is devoted to van der Waals epitaxy of 
both planar structures \cite{hong2015van, utama2013recent} and NWs’ arrays, 
including ZnO \cite{KoreaPaper}, Zn$_3$P$_2$ \cite{paul2020van},
GaAs \cite{munshi2012vertically,Berd2019,alaskar2014towards} , InAs \cite{Hong2011} and 
GaN \cite{MNL, CGD, sundaram2019large}.
Note, that graphene seeding layers attract a lot of attention for 
III-Nitrides NWs growth due to the hexagonal arrangement of the carbon 
atoms with sp$^2$ hybridization, matched to the (0001) c-plane of wurtzite GaN 
\cite{al2015impact}. 
Several groups reported the synthesis of GaN NWs and column structures 
on pristine or defective graphene flakes \cite{MorassiNWs,fernandez2017molecular}. 
However, the detailed theoretical analysis the GaN NW nucleation on 
virtual graphene substrates has not been reported so far.

Due to inversion asymmetry along the c-axis, wurtzite GaN nanostructures may 
form N- or Ga-terminated facets corresponding to different polarity. 
The specific facet polarity of nitride NWs impacts their electronic 
\cite{stutzmann2001playing} and nonlinear optical properties \cite{hite2012development}, 
and therefore, affects the design of piezoelectric \cite{gogneau2016single}
and light-emitting \cite{carnevale2013mixed} applications. 
N-polar surfaces were reported to have the advantages of low-resistivity ohmic 
contacts and improved capability for large-scale processing for 
high electron mobility transistors \cite{wong2013n}.
GaN NWs grown by molecular beam epitaxy on conventional 
substrates commonly exhibit N-polarity \cite{hestroffer2011polarity}, 
however the question of the polarity of GaN NWs grown on graphene has not been addressed so far. 

In this work, we employ the density functional theory (DFT) based analysis to study 
and explain the experimentally observed polarity discrimination in GaN nanoislands 
and NWs epitaxially grown on graphene.
%
The Kohn-Sham formalism \cite{kohn1965self, hohenberg1964inhomogeneous} 
of DFT has proved itself as one of the most indispensable methods for 
electronic structure calculations.
%
%
Nowadays, DFT is routinely used to predict equilibrium properties of the molecules, 
bulk materials and nanostructures \cite{sholl2011density, dobson2013electronic}.
However, our work, for the first time to our knowledge, 
addresses the mechanism of polarity discrimination within the study 
of  the preferred structure and polarity of GaN nanoclusters.
%
%
%
%
We tentatively explain this preference  of N-polarity as being  caused by dipole 
interaction between the GaN nanocrystal and $\pi$-orbitals of the graphene. 
Theoretical results are 
fully agreed with the experimental observations on
self-induced vertically-oriented GaN NWs grown on CVD-graphene 
by plasma assisted molecular beam epitaxy (PA-MBE).
%

\section{Experimental results}

\begin{figure*}[ht]
  \centering
  \includegraphics[width=\textwidth]{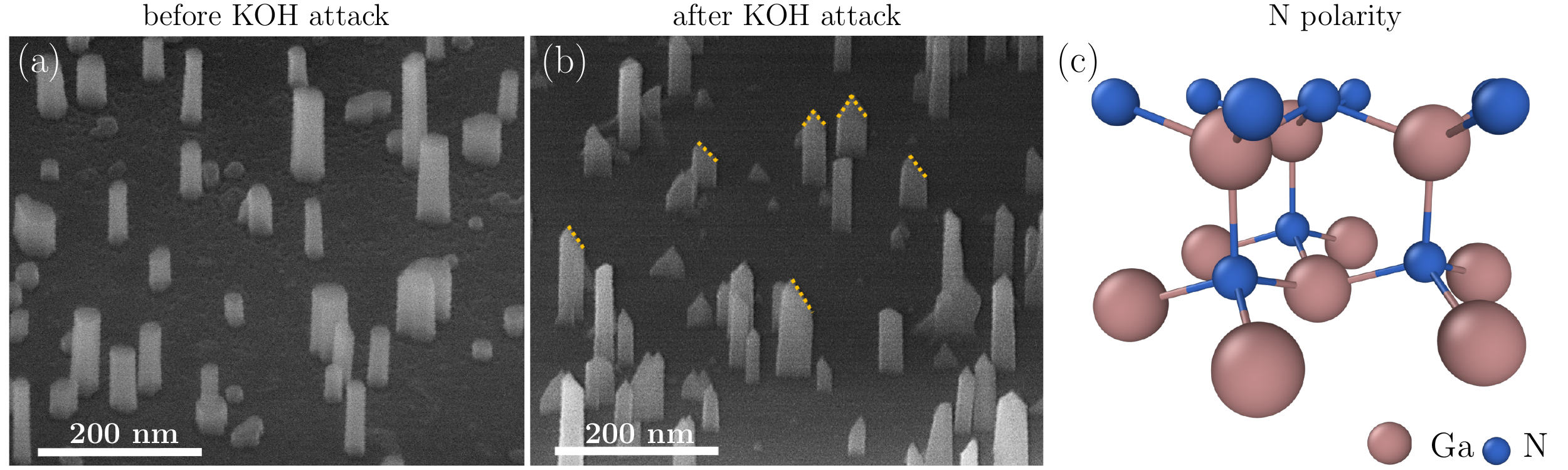} 
  \caption{(a) 45° tilted SEM image of as-grown NWs on a graphene patch,
           (b) SEM image of the same sample after KOH attack.
           Yellow lines highlight \textit{exemplary} pencil-like top shapes.
           Schematic illustration of (c) a N-polar GaN  wurtzite crystal.
           Brown spheres show the position of gallium atoms and 
           blue ones show the position of nitrogen atoms. 
           Lines are corresponding chemical bonds.}  
  \label{fig:sem_pol}
\end{figure*}

Many groups have demonstrated synthesis of GaN NWs on the graphene by different growth techniques 
\cite{heilmann2016vertically, kumaresan2016epitaxy, fernandez2017molecular, lee2011flexible}. 
In particular, in our previous experimental study we reported the growth of GaN NWs 
on graphene transferred to an amorphous host substrate \cite{MNL, CGD}. 
We used commercially available CVD polycrystalline graphene grown on a Cu foil 
transferred to a Si/SiO$_2$ template. 
It was demonstrated that vertical NWs can be selectively grown on graphene patches 
and that the in-plane NW orientation is imposed by the graphene layer 
in a way that the $\langle$2-1-10$\rangle$ directions of the wurtzite 
GaN lattice are parallel to the directions of the zigzag chains of 
the honeycomb carbon lattice \cite{kumaresan2016epitaxy}. 

For the present investigation, we adopted the growth conditions yielding high 
growth selectivity on the graphene with respect to the SiO$_2$ surface \cite{CGD}.
Graphene patches of 1 cm$^2$ were wet-transferred to Si(100) 
substrates topped with a 300 nm layer of thermal oxide. 
The substrate was outgassed in the growth chamber at 830$^{\circ}$C during 5 min, 
then the temperature was stabilized at 815$^{\circ}$C. 
Ga and N fluxes were provided simultaneously. 
The V/III ratio was equal to 1.1 and the Ga flux was equivalent to a 2D GaN growth rate 
of 0.7 monolayers per second. The NW nucleation was monitored by reflection
high-energy electron diffraction. 
The first diffraction spots corresponding to the formation of GaN nuclei were observed 
after 90 min time of exposure to the fluxes and then the growth was 
continued for additional 40 min yielding the total growth time of 130 min. 
Figure \ref{fig:sem_pol} (c) shows a scanning electron microscopy 
(SEM) image of the resulting NW morphology.
The NW density is about 1.8$\times$10$^{10}$ cm$^{-2}$, the average 
diameter and height are 25 nm and 160 nm, respectively. 

It is a common result for different types of substrates, 
that MBE-grown GaN NWs exhibit N-polarity \cite{kumaresan2016self,de2012polarity}. 
Yet the polarity was not investigated for the specific case of NWs 
on graphene/SiO$_2$ templates. 
To determine the polarity in thin NWs, potassium hydroxide (KOH) etching 
is a simple and effective method 
\cite{hestroffer2011polarity, largeau2012n}.
Indeed, KOH selectively etches N-polar surface by forming hexagonal pyramids, 
while Ga polar surface remains unaltered. 
This selectivity is associated with the surface bonds configuration 
for N- and  Ga-polar surfaces \cite{li2001selective}, 
which is schematically illustrated in Figures \ref{fig:sem_pol} (c)
respectively: on Ga polar surface, OH$^-$ ions from KOH are strongly repelled 
by the electronegativity of the three N dangling bonds, while for N polar surface, 
OH- ions are adsorbed on the surface to form Ga$_2$O$_3$ by reacting with 
Ga which later dissolves in the KOH solution. 

To evaluate the polarity of GaN NWs grown on graphene, we used a KOH solution 
of concentration of 0.5 mol/l to etch the NWs for 2 minutes at 40 $^{\circ}$C and, 
then the sample was observed in SEM as shown in Figure \ref{fig:sem_pol} (b). 
By comparing Figures \ref{fig:sem_pol} (a) and (b) displaying the images before and after 
the KOH attack it can be seen, that all the as-grown NWs exhibit a flat top, 
while after the KOH attack the top shape becomes pencil-like. 
This transformation is a typical signature of N-polar NWs 
\cite{hestroffer2011polarity, largeau2012n}, 
which confirms that GaN NWs grown on graphene present the same polarity 
as NWs grown on other substrates.

\section{Computational analysis}

In order to find energetically favorable polarity of the NW, 
we computationally analyse the equilibrium configuration for
the small GaN nanocrystals (NCs) on a graphene sheet, 
which are the precursors for further GaN NWs growth.

Let us introduce the system under consideration. 
Firstly, in our calculations only the graphene sheet was considered as the substrate. 
Graphene is commonly considered to interact with silicon 
oxide layer through weak van der Waals forces \cite{Fan_2012}
Indeed, as shown in experimental work \cite{CGD} a graphene flake transferred 
to a relatively thick oxide layer substrate does not interact 
strongly with the substrate underneath. 
This is evidenced by the thermal expansion of the graphene 
flake independent from the thermal expansion of the Si substrate supporting the flake. 
In this case the graphene thermal expansion coefficient was found to be several 
times lower compared to Si \cite{CGD}. 
Thus, in our calculations we can neglect the GaN interaction with SiO$_2$ .

Next, we define the initial relative position of Ga and N atoms 
in the NC with respect to the graphene lattice.
%
%
As proposed in \cite{MNL, CGD}, (3$\times$3) GaN and 
(4$\times$4) graphene supercells adopt the in-plane alignment of GaN and graphene lattices. 

\begin{figure}[ht]
  \centering
  \includegraphics[width=0.5\textwidth]{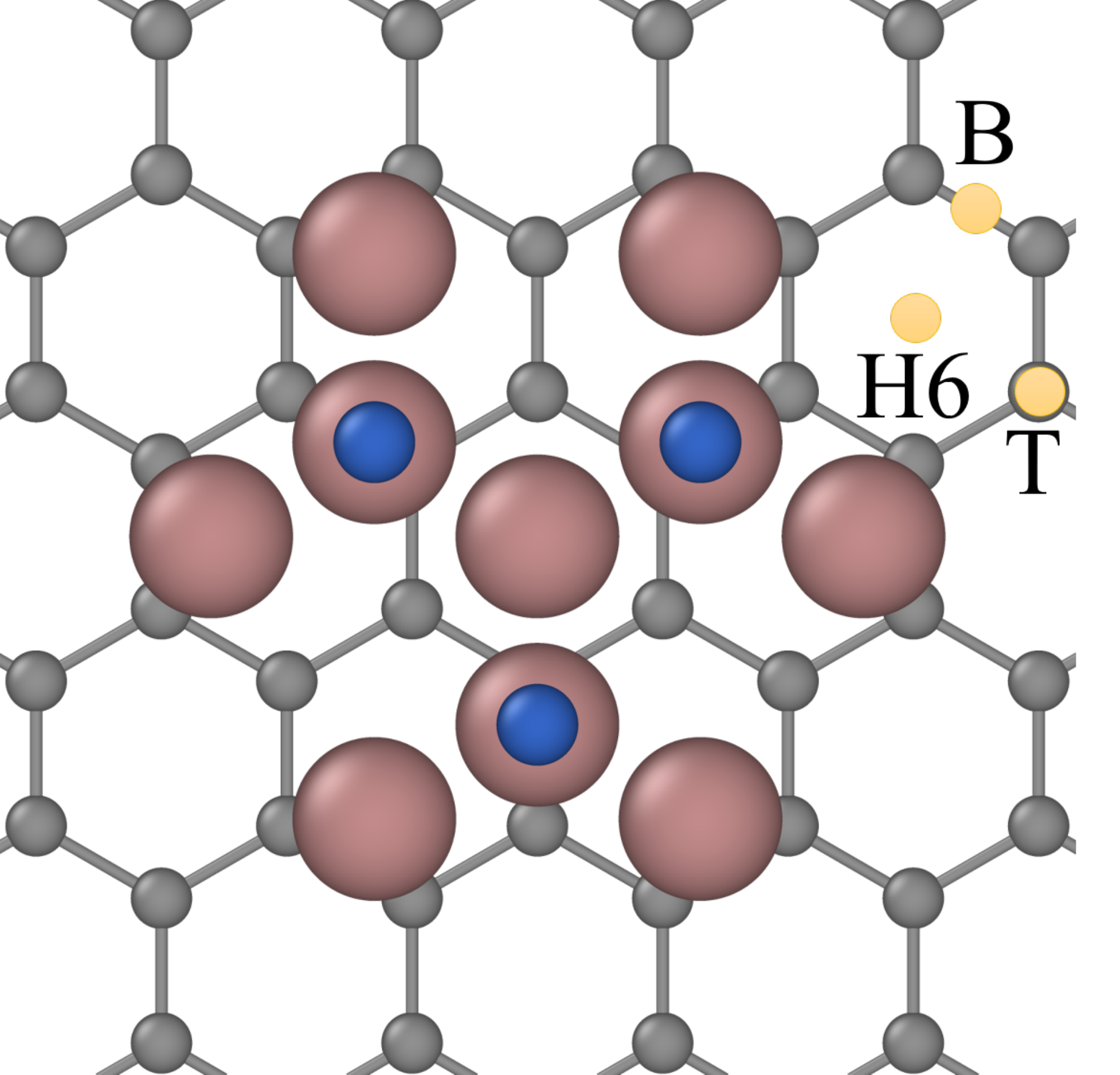}
  \caption{\textit{Exemplary} top view of the system considered. 
            The $\langle 2-1-10 \rangle$ direction of the GaN cluster 
            was chosen to be aligned to the zigzag direction of graphene sheet.
            Labels H6, T and B indicate the "hollow", "top" and "bridge"
            positions on the graphene cell respectively.}
  \label{fig:top_view}
\end{figure}

To verify this assumption, we investigated the energy preferable positions of 
individual Ga and N atoms on graphene sheet. 
Similar to previous works \cite{munshi2012vertically,Nakada2011}, 
we analyzed the adsorption of Ga and N atoms at three types of sites 
("top", "bridge" and "hollow") shown in Figure \ref{fig:top_view}.  
For nitrogen atoms the "bridge" and "top" sites were found 
to be energetically favorable. 
While for gallium atoms all the three sites have shown very similar binding energies. 
The detailed discussion is given in Supplementary materials, section 1.

Therefore the reported \cite{MNL} matching between (3$\times$3) GaN and 
(4$\times$4) graphene supercells corresponds to the minimal binding 
energy of Ga and N atoms. 
In this case the $\langle$2-1-10$\rangle$ 
directions of the wurtzite GaN lattice are parallel to the directions 
of the zigzag chains of the honeycomb carbon lattice. 
To reflect this lattice symmetry the NC of at least ten Ga and ten N atoms 
(Ga$_{10}$N$_{10}$) should be considered. 
Figure \ref{fig:top_view} gives an example of the top view of Ga$_{10}$N$_{10}$
NC in the initial configuration. 
%

%
%
It is crucial that the chosen size of the cluster is sufficient to reconstruct 
the symmetry of the wurtzite GaN on graphene. 
So one can translate such NC to reconstruct the arbitrary large NW. 
Meantime our approach also requires the considered GaN NC to be 
larger than the size of the critical nucleus for a GaN nanoisland.
Previous reports on self-induced nucleation of GaN nanostructures 
typically estimated the critical nucleus to a few III-V pairs 
\cite{sobanska2016analysis, consonni2011physical, sobanska2019comprehensive}.
Such a small critical nucleus cannot be observed experimentally and 
gives the ground to the assumption of the irreversible 
growth of catalyst-free NWs on silicon substrates with 
Si$_{\rm x}$N$_{\rm y}$ and Al$_{\rm x}$O$_{\rm y}$ 
amorphous layers \cite{sobanska2016analysis, fernandez2015monitoring}.
In Supplementary material we show that those typical assumptions of small 
critical nucleus and irreversible growth are also reasonable 
for the considered in this work catalyst-free GaN NCs on graphene. 
Therefore, the size of the GaN NC considered in our model is not limited 
by the size of the critical nucleus, though it should satisfy 
the domain matching between graphene and GaN lattices.

\begin{figure*}[ht]
    \centering
    \includegraphics[width=\textwidth]{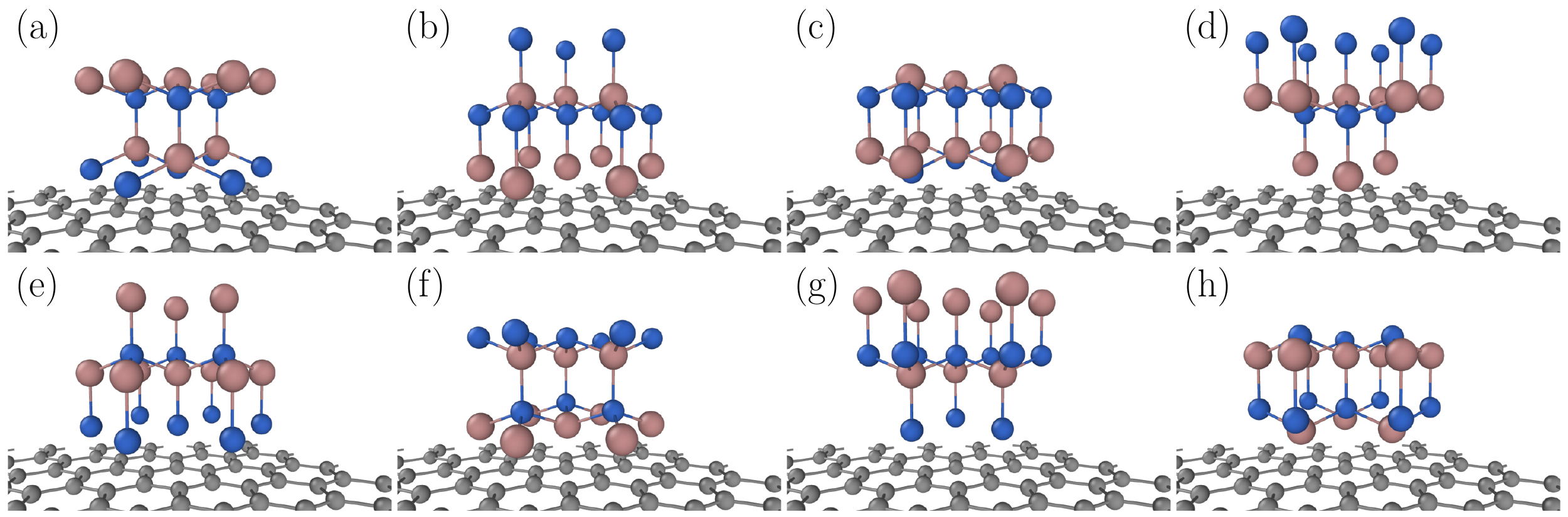}
    \caption{Examples of all possible crystalline combinations 
              in case of Ga$_{10}$N$_{10}$ nanocrystals. 
              For both polarities it is possible to construct nanocrystals
              which start from layers (the closest to graphene layers) 
              of 7 N, 7 Ga, 3 N or 3 Ga atoms respectively.
              First row (a) - (d) shows the Ga polar nanocrystals,
              second row (c) - (h) shows the N polar nanocrystals.}
    \label{fig:initial_configurations}
\end{figure*}

There are 8 possible configurations of the Ga$_10$N$_10$ 
NCs which are all shown in Figure \ref{fig:initial_configurations}: 
four configurations are Ga-polar (upper row) and the other four are N-polar (lower row).
%
%
To reconstruct the wurtzite structure each hexagonal Ga$_{10}$N$_{10}$ 
NC should have four layers: two layers of seven 
and two of three gallium or nitrogen atoms, labeled as 7Ga, 7N, 3Ga and 3N, respectively.
The layer closest to graphene determines the initial configuration 
while the polarity of a crystal determines the order of remained layers.
%
%
For example in Figure \ref{fig:initial_configurations} (a)
the NC consisting of 7 N - 3 Ga - 3 N - 7 Ga layers is depicted.

Further, the full self-consistent optimization of the geometry of each systems 
was performed by \textit{ab-initio} approach. 
The \textit{ab-initio} calculations were carried out 
using plane-wave basis set for projector augmented-wave method
implemented in GPAW package \cite{gpaw}. 
The starting spatial structure of the systems under consideration 
was made by ASE library \cite{ase}.
The simulation cell had periodic boundary conditions 
and sizes 14.70 $\times$ 12.25 $\times$ 14.00 \AA$^3$.
The total number of atoms in the calculation was 80, 
where 60 of them were carbon atoms forming graphene sheet.
%
%
To prevent interaction between NCs though the periodic 
boundary the lateral distances between NCs was $\approx$ 6 \AA\
and the vacuum gap along $z$ direction was $\approx$ 8 \AA. 
The full optimization of the spatial structure within DFT for the relaxation 
of a stressed system and determination of the equilibrium configuration 
was done by using the modified 
Broyden – Fletcher – Goldfarb – Shanno algorithm \cite{nocedal2006numerical} 
implemented in GPAW \cite{gpaw}. 
The optimization process ends when mean force acting on the atoms become 
less than a cutoff value (0.05 eV/\AA). 
%
%
The graphene atoms were fixed during optimization, which allowed to
reduce the number of optimization steps without significant losses 
in prediction accuracy.  

For each step of geometry optimization the self-consistent calculation of the electronic 
density was made with the use of the Perdew–Burke–Ernzerhof (PBE) generalized 
gradient approximation (GGA) exchange-correlation functional. 
The Monkhrost-Pack grid 4$\times$4$\times$4 was used in calculations 
and the plane-wave cutoff was set to $450$ eV. 
Calculations were performed with convergence threshold 
for self consistency of the charge density $10^{-4}$.

We found that only two initial geometries,
(Ga-polar 7 N - 3 Ga - 3 N - 7 Ga  Figure \ref{fig:initial_configurations} (a)
and N-polar 3 Ga - 7 N - 7 Ga - 3 N  
Figure \ref{fig:initial_configurations} (h) ) remained stable after the optimisation procedure.
In the other systems it was more energetically favorable for nitrogen to form the
N$_2$ molecule and "evaporate" from the NC.
This nitrogen "evaporation" effect is common for calculations involving nitrogen 
and special attention should be paid to account for it \cite{de1998structural}.
The snapshots of the system geometries during optimization are shown 
in Supplementary material section 2. 
%
with finite size of the systems considered.
%
%
%
%

\begin{figure*}[ht]
  \centering
  \includegraphics[width=\textwidth]{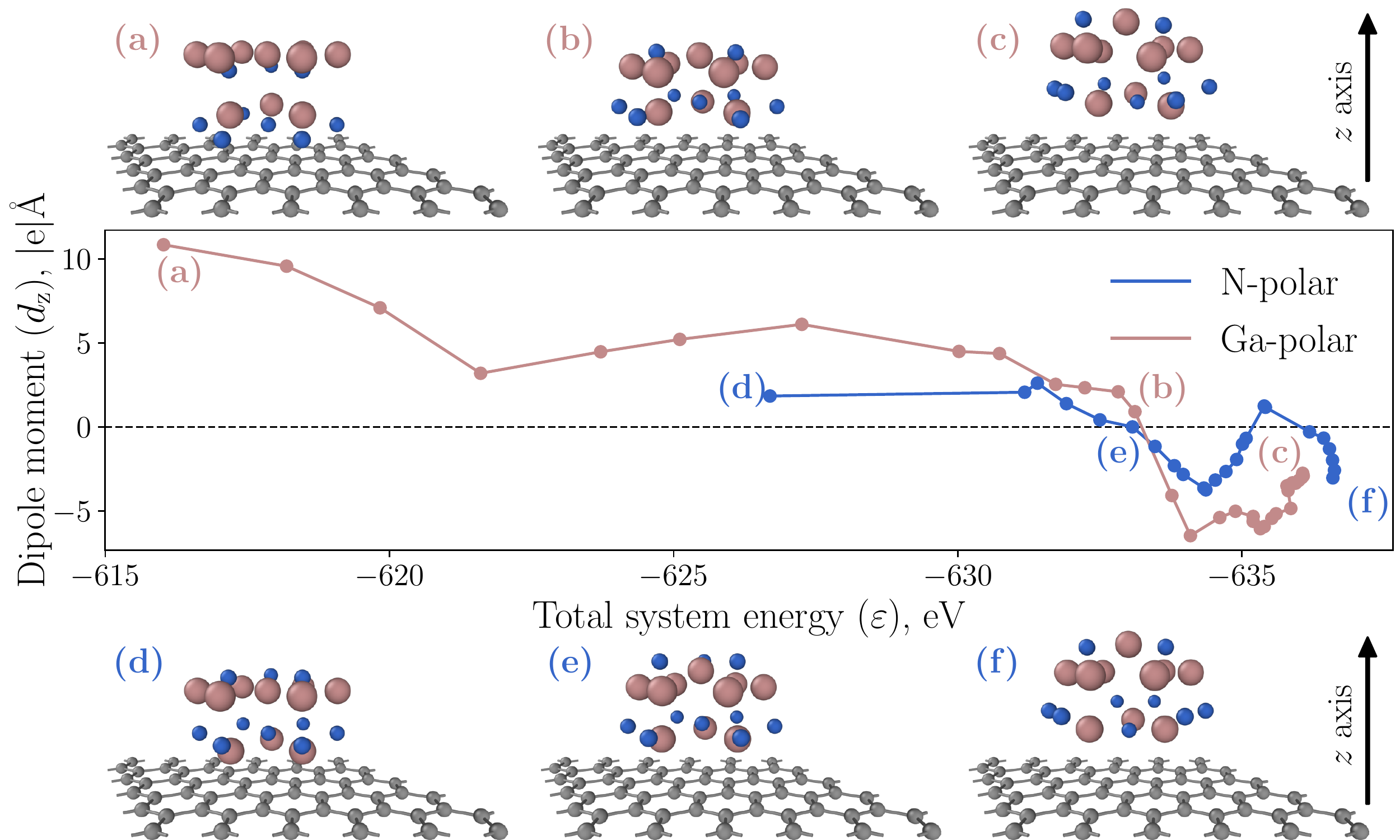}
  \caption{The snapshots of geometries of 
            NCs during optimization procedures.
            Grey spheres and lines correspond to carbon atoms and bonds in graphene. 
            Brown ones to gallium and blue ones to nitrogen atoms.
            (a) and (d) the initial configurations, 
            (b) and (e) the intermediate configurations and (c) and (f) final, 
            optimized structures of Ga and N-polar NCs, respectively. 
            Central panel shows the optimization trajectories 
            of Ga and N polar NC plotted in the dipole moment vs total system energy 
            ($d_{\rm z}$ vs $\varepsilon$) coordinates. 
            Latin letters correspond to the system configurations shown on the figures
            above and below. }
  \label{fig:dipole_moment}
\end{figure*}

Let us now analyse the stable configurations of the NCs. 
%
%
Figure \ref{fig:dipole_moment} shows the snapshots of the system geometry during the optimization
for Ga-polar (panels (a) - (c)) and N-polar (panels (d) - (f)) NCs.
It is seen, that the initial configurations of these systems have different orientation 
of the Ga-N dipole.
For Ga-polar NC the initial configuration (Figure \ref{fig:dipole_moment} (a)) 
has the following order of atomic layers 7N - 3Ga - 3N - 7Ga.
The intermediate state is shown in Figure \ref{fig:dipole_moment} (b), 
in that state 7N - 3Ga and 3N - 7Ga planes are almost at the same level.
At the final optimization step shown in Figure  \ref{fig:dipole_moment} (c) 
the position order of atomic planes along the growth direction 
has changed to 3Ga - 7N - 7Ga - 3N. 
It is worth to mention, that during the optimization procedure NCs maintain
the initial crystalline orientation with respect to the graphene sheet,
only a rearrangements along the $z$ axis was observed.
So, during the optimization procedure Ga-polar NC 
changes the orientation of its Ga-N dipole, i.e. the polarity.
On the other hand, the N-polar NC only adjusts its shape, 
without any significant rearrangement of the atomic planes 
and changes in the Ga-N dipole orientation.
%
%

As a result, GaN NCs which had different polarities before optimization, 
after optimization procedure have the same polarities 
and resembling geometrical structures.
To further analyze the evolution of the GaN NCs 
during the optimization one needs to define a physical quantity that characterizes
the polarity of the system.

The difference between Ga- and N-polarity structures  
can be described by projection of the electric dipole moment 
of the system on the $z$ axis.   
The dipole moment $d$ of the system can be represented as
a sum of atomic dipole moments $d^{\rm at}$ and dipole moments 
due to net charges which occur because of a non-uniform 
distribution of the electron density of the system. 
So the z-projection of the dipole moment $d_z$ can be written as:

\begin{equation}
    d_z = \sum_{i=1}^{N_{\rm at}} \left( d^{\rm at}_{i} + q_{i} r_{i}\right ),
    \label{eq:dipole_moment}
\end{equation}
where $r_{i}$ - the atomic positions, $q_i$ - effective electric net charges 
corresponding to each atom with taking into account of 
the valence electron charge density distribution and 
$N_{at}$ - number of atoms of GaN system. 
The values of the net charges are determined by the 
charge transfer between different atoms into the system and 
can be obtained by Bader analysis \cite{bader1, bader2, bader3}. 
The atomic dipole moments were calculated through atomic 
polarizability and effective electric field corresponding 
to the current charge distribution. 

The central panel of Figure \ref{fig:dipole_moment} shows the dependence 
of the dipole moment of the system 
along the $z$ axis on the total energy of the system for the two stable configurations.
One can notice that the initial configurations of Ga and N-polar systems 
(Figure \ref{fig:dipole_moment} (a) and (d), respectively) 
have very different positions on the plot.
N-polar system has a significantly lower total energy and dipole moment.
Despite different initial configurations both systems  
in optimized state have the same dipole moment and similar total system energies. 
%
%
The shape of the optimization curve depends strongly on the optimization 
algorithm used, but the final geometries of the structures are the same.
%
%
Both optimization curves cross the zero level of dipole moment 
(Figure \ref{fig:dipole_moment} (b) and (e)) and 
final structures correspond to negative $d_z$ values
(Figure \ref{fig:dipole_moment} (c) and (f)).
%
%

\begin{figure}[ht]
  \centering
  \includegraphics[width=0.5\textwidth]{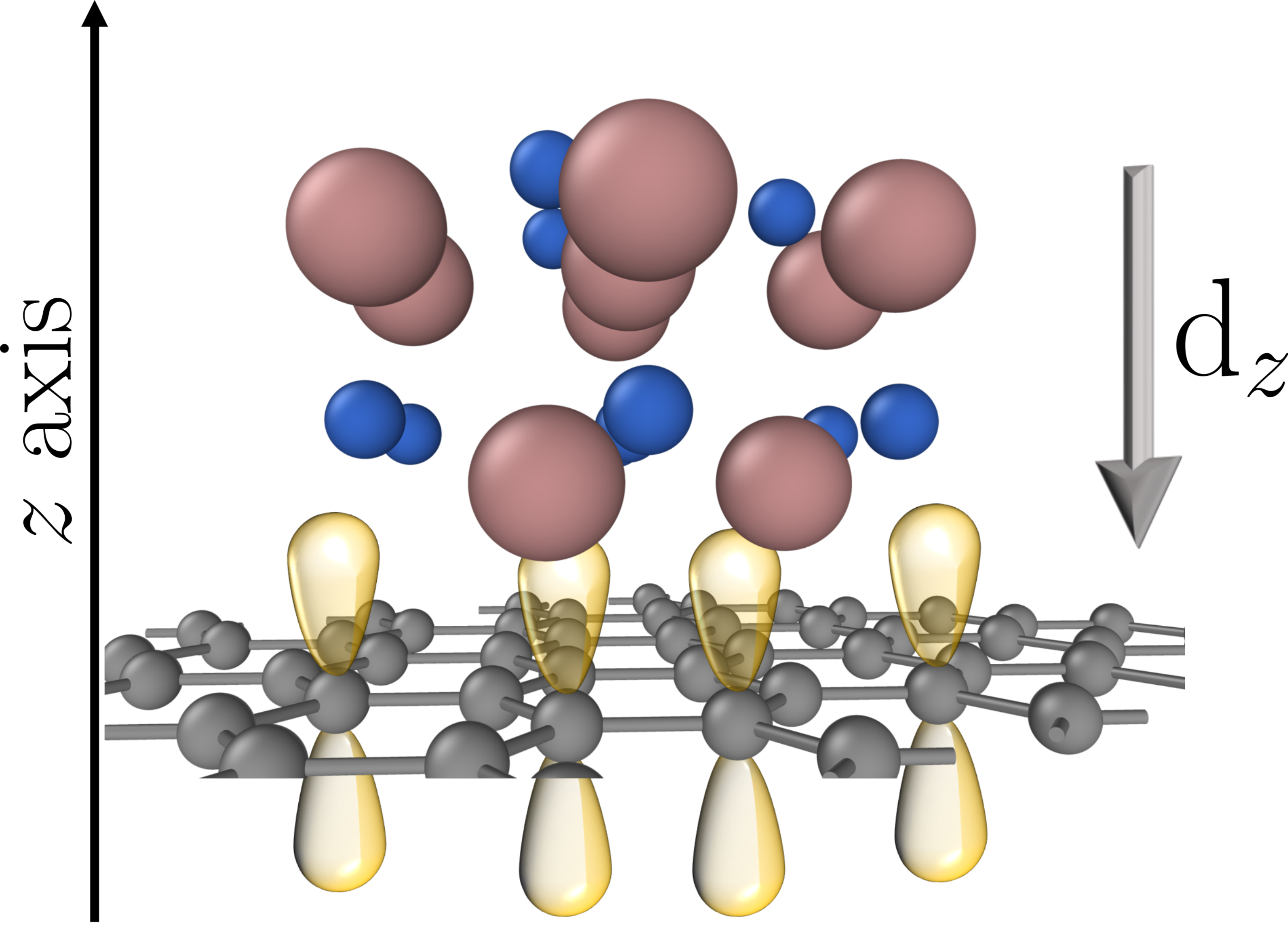}
  \caption{Sketch of the GaN nanocrystal placed on top of graphene sheet.
           The bold arrow shows the direction of the nanocrystal dipole moment.  
           Yellow areas illustrate the position and shape of graphene $\pi$-bonds (not in scale).
           z-axis is placed along the NW growth direction.
           }
\label{fig:pi-bounds}
\end{figure}

Thus, experimental and numerical studies revealed the polarity discrimination 
which can be explained by GaN interaction with graphene charge density. 
The electron density of graphene extended along the $z$ -axis forms negative 
charge density localized in the space between the GaN NC and the 
graphene as illustrated in Figure \ref{fig:pi-bounds}. 
Therefore, the electric dipole moment of the GaN NC tends to turn 
towards the negative charge, which results in a negative dipole moment projection $d_z$. 
Therefore, the lowest energy configuration has the electric dipole moment of 
the GaN NC turned towards the negative charges. This makes the N-polar 
configuration energetically stable while Ga-polar 
configurations tend to switch the polarity.

%
%

\section{Conclusions}

In this work we have demonstrated experimentally the polarity discrimination 
for GaN NWs MBE grown on graphene and explained the observations theoretically.
%
We have presented the DFT studies of a GaN cluster with a supercell matching 
to graphene and a size that exceeds the critical nucleus for a GaN nanoisland. 
We have considered all possible configurations of Ga$_{10}$N$_{10}$ 
NCs on graphene and have shown that only two of them 
(Ga-polar starting from layer of 7 nitrogen atoms  and 
N-polar starting from layer of 3 gallium atoms 
Figure~\ref{fig:initial_configurations} (a) and (h) respectively) are stable,
while others tend to dissociate.

The DFT energy optimization shows that the considered Ga-polar NC changes 
the order of atomic layers turning into N-polar configuration. 
We introduced the projection of the electric dipole moment of the system 
as the physical quantity characterizing the NC evolution during the optimization.
%
Indeed, for both NCs the dipole moment projection is negative at minimum 
energy configurations which corresponds to N-polarity.

Our results show that the observed phenomenon can be explained 
by considering the interaction of the electric dipole of the NCs 
with net charges formed by $\pi$-orbitals of the graphene sheet. 
The stable GaN NC with the electric dipole moment turned towards 
the graphene negative charges is preferred due to the lowest system energy. 
Thus, the GaN nanostructures tend to show N-polarity.


These results can be applied for engineering of GaN NW-based optoelectronic 
devices and crucial for design of piezogenerators where the opposite polarity of 
different NWs can reduce the efficiency of pressure-to-voltage conversion.

\begin{acknowledgments}
  The work was financially supported by the 
  Ministry of Science and Higher Education 
  of the Russian Federation (FSRM-2020-0005).
  This work was done with the financial support of 
  the Russian Federation President Council 
  for grants (grants MK-2428.2020.2 and SP-2169.2021.1).
  Y.B. acknowledges the support of growth modeling by 
  the Russian Science Foundation under the Grant 19-72-30004
  C.B. and M.T. acknowledge the financial support from 
  the EU ERC project NanoHarvest (grant no. 639052) 
  and Labex “GaNeX” (ANR-11- LABX-2014).
  We acknowledge the Supercomputing Center of 
  Peter the Great Saint-Petersburg Polytechnic University 
  (SPbPU) for providing the opportunities to 
  carry out large-scale simulations. 
  A.P. is grateful to Alexander Ustinov (SPbPU) for fruitful discussion.
  The molecular structures in this article 
  were rendered using Ovito software package \cite{ovito}. 
\end{acknowledgments}

\bibliography{gan}

\end{document}